\documentclass[twocolumn,prl,superscriptaddress,showpacs,floats]{revtex4}
\usepackage[pdftex]{graphicx}
\usepackage{dcolumn}
\usepackage{color}

\usepackage{multirow}

\usepackage{bm}

\begin{document}

\title{Giant magnetic anisotropy of Co, Ru, and Os adatoms on MgO (001) surface}
\author{Xuedong Ou}
\affiliation{Laboratory for Computational Physical Sciences (MOE), State Key Laboratory of Surface Physics, and Department of Physics, Fudan University, Shanghai 200433, China}
\affiliation{Science and Technology on Reliability and Environmental Engineering Laboratory, Beijing Institute of Spacecraft Environment Engineering, Beijing 100094, China}
\author{Hongbo Wang}
\affiliation{Laboratory for Computational Physical Sciences (MOE), State Key Laboratory of Surface Physics, and Department of Physics, Fudan University, Shanghai 200433, China}
\author{Fengren Fan}
\affiliation{Laboratory for Computational Physical Sciences (MOE), State Key Laboratory of Surface Physics, and Department of Physics, Fudan University, Shanghai 200433, China}
\author{Zhengwei Li}
\affiliation{Laboratory for Computational Physical Sciences (MOE), State Key Laboratory of Surface Physics, and Department of Physics, Fudan University, Shanghai 200433, China}
\author{Hua Wu}
\thanks{Corresponding author: wuh@fudan.edu.cn}
\affiliation{Laboratory for Computational Physical Sciences (MOE), State Key Laboratory of Surface Physics, and Department of Physics, Fudan University, Shanghai 200433, China}
\affiliation{Collaborative Innovation Center of Advanced Microstructures, Fudan University, Shanghai 200433, China}

\date{today}

\begin{abstract}

Large magnetic anisotropy energy (MAE) is desirable and critical for nanoscale magnetic devices. Here, using ligand-field level diagrams and density functional calculations, we well explain the very recent discovery [I. G. Rau et al., Science 344, 988 (2014)] that an individual Co adatom on a MgO (001) surface has a large MAE of more than 60 meV. More importantly, we predict that a giant MAE up to 110 meV could be realized for Ru adatoms on MgO (001), and even more for the Os adatoms (208 meV). This is a joint effect of the special ligand field, orbital multiplet, and significant spin-orbit interaction, in the intermediate-spin state of the Ru or Os adatoms on top of the surface oxygens. The giant MAE could provide a route to atomic scale memory.

\end{abstract}


\maketitle

Surface-embedded molecular magnet structures are of great interest, as they have the potential for miniaturizing magnetic units at the ultimate atomic scale.\cite{Kha11,Kha13,Loth} Recently, studies about magnetic adatoms with a large MAE are extremely vigorous for their promising applications in high density information storage and quantum spin processing.\cite{Gam,Hir,Blonski,Donati,Hu,Bel,Xiao,Rau,Kha14} For example, single Co atoms deposited onto a Pt (111) surface give rise to a MAE of 9 meV per atom, and the assembled Co nanoparticles have a single-atom coordination dependent MAE.\cite{Gam} In addition, those large-MAE systems include Fe or Mn atoms absorbed on the CuN surface,\cite{Hir} Co and Fe atoms on Pd and Rh (111) surfaces,\cite{Blonski} and Co or Ir related dimers on surfaces of graphene, defected graphene or benzene as well.\cite{Donati,Hu,Bel,Xiao} A giant MAE would produce an energy barrier to protect a stable and robust magnetization from thermal fluctuations, thus enabling the magnetization to be oriented along a preferential spatial direction for a sufficient amount of time.

Strategies to enhancing the MAE of magnetic adatoms consist of three important factors --- a giant spin-orbit coupling (SOC) energy, a special ligand field, and a large orbital moment.\cite{Rau,Kha14} As a ligand field often quenches or diminishes an orbital moment, searching for a suitable surface or substrate is a challenge for achieving a giant MAE. Very recently, Rau et al. found, by constructing an optimal strategy, that a huge MAE of about 60 meV has been achieved for the Co atoms adsorbed on top of the O sites of MgO (001) surface.\cite{Rau,note} This MAE is record breaking and reaches the magnetic anisotropy limit of a 3d metal atom. Enhancing the magnetic properties of adatoms provides a route to atom-scale memory.\cite{Kha14}

In this Letter, we first provide an insight into the above discovery through ligand-field level analysis and density functional calculations. We have sought the origin of the huge MAE for Co adatoms on MgO (001), by addressing the orbital multiplet effect of the high-spin Co adatom in the special tetragonal ligand field. More importantly, we demonstrate that Ru or Os adatoms on MgO (001) also reside stably on top of the O sites and both remain in an intermediate-spin state. Then their significant SOC, together with the orbital multiplet effect, produces a giant MAE of 110 meV/Ru and 208 meV/Os both with an easy out-of-plane magnetization. In this sense, magnetic anisotropies would be engineered extremely high to produce stable magnetization at room temperature for a single atomic spin.

\begin{figure}[b]
\centering \includegraphics[width=7cm]{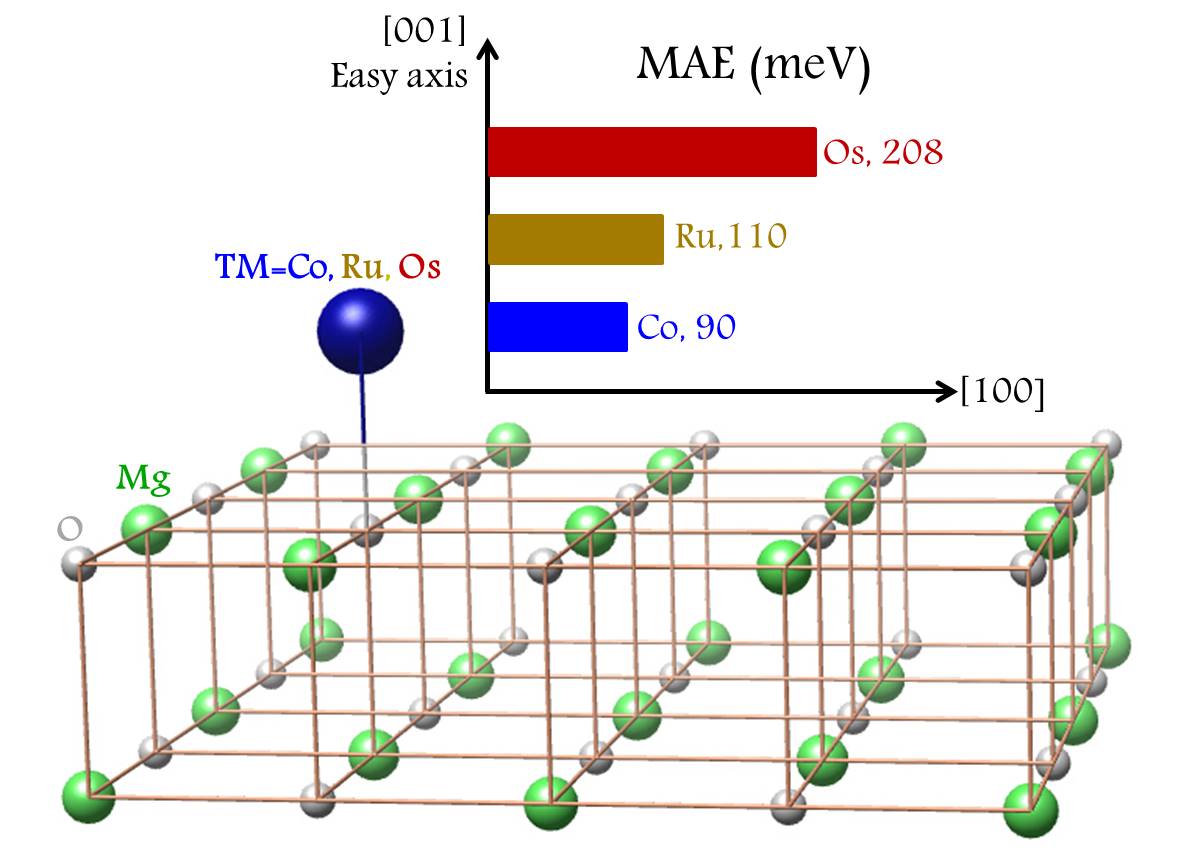}
\caption{(Color online)
Stable O-top adsorption site of the TM adatoms on MgO (001). The MAEs are also illustrated.
}
\end{figure}

We have used a slab model with a 10-\AA vacuum to simulate the transition-metal (TM) adsorbed MgO (001) surface. One monolayer of the highly insulating MgO is chosen, and a lateral $2\sqrt{2}\times2\sqrt{2}$ supercell is adopted, with a single TM adatom. All the atoms are relaxed along the surface normal direction till each atomic force is smaller than 5 meV/\AA. We have used the full-potential augmented plane wave plus local orbital code (Wien2k).\cite{Blaha} The muffin-tin sphere radii are chosen to be 2.1, 2.0 and 1.2 Bohr for TM (Co, Ru, Os), Mg and O atoms, respectively. The cutoff energy of 12 Ryd was used for the interstitial plane wave expansion and $12\times12\times1$ k mesh for integration over the Brillouin zone. We employ the local spin density approximation plus Hubbard U (LSDA+U) method\cite{Anisimov} to describe the electron correlation of the TM adatoms. The typical values of the Hubbard U (and Hund exchange J), U = 6 eV (J = 1 eV) are used for Co 3d electrons,\cite{Wu09} 3 eV (0.5 eV) for Ru 4d,\cite{Wu06} and 2 eV (0.4 eV) for Os 5d.\cite{Wang} The change $\Delta$U = 1 eV brings about a only few-meV change for the calculated giant MAE and hence does not affect our conclusion at all. The SOC of those valence d electrons is included by the second-variational method with scalar relativistic wave functions.\cite{Blaha} The MAE is calculated through the total energy difference for two magnetization directions (out of plane versus in plane).

We first search the stable adsorption sites of the Co (Ru and Os as well) adatoms on MgO (001) surface (see Figure 1), using LSDA total energy calculations including atomic relaxation. Our results show that the O-top site is energetically most favorable, being lower in total energy than the Mg-top site and the O-O or Mg-Mg bridge (i.e., hollow) site by 1.46 eV and 0.78 eV per Co adatom, respectively. The corresponding results for Ru and Os adatoms are summarized in Table I. This finding of the most stable O-top site agrees with the recent experimental observation for Co/MgO (001).\cite{Rau}

\begin{table}[b]
\caption{Relative energies (in unit of eV) for TM adatoms on MgO (001) at three different adsorption sites. The O-top site is most stable, and the corresponding MAE (meV), spin and orbital moments ($\mu_B$) along the easy [001] axis are listed.}
\begin{tabular}{l@{\hskip5mm}c@{\hskip5mm}c@{\hskip5mm}c@{\hskip3mm}c
@{\hskip3mm}c@{\hskip3mm}c}
\hline\hline
 TM & Mg-top & Bridge & {\bf O-top} & {\bf MAE} & M$_{spin}$ & M$_{orb}$  \\ \hline
 Co & 1.46 & 0.78 & 0 & 90 & 2.44 & 2.48 \\
 Ru & 1.39 & 0.61 & 0 & 110 & 1.73 & 1.81 \\
 Os & 1.56 & 0.91 & 0 & 208 & 1.46 & 1.70 \\
\hline\hline
\end{tabular}
\end{table}

At the most stable O-top adsorption site, the Co 3d electrons see a special tetragonal ligand field, and the schematic level diagram is plotted in Figure 2(a). Owing to the underlying O$^{2-}$ ion, both the ionic and covalent contributions of the ligand field raise the $3z^2-r^2$ electronic level to the topmost one, which is followed by the two-fold $xz$/$yz$ levels. The planar $xy$ and $x^2-y^2$ orbitals lie at the lowest energy, and they are almost degenerate, with the $x^2-y^2$ level being slightly lower due to the attractive interaction with the next nearest neighboring Mg$^{2+}$ cations. Apparently, for such a low-coordinated Co adatom sitting at the O-top site, the ligand field effect is not as significant as in a TM-O polyhedron (e.g., a common CoO$_6$ octahedron). As a result, the Hund exchange dominates, and the Co adatom has a high-spin (HS) state with S=3/2 for the $3d^7$ configuration, see Figure 2(a).

When one has a glance at the crystal field level diagram of the HS Co adatom, one may assume that the two minority-spin electrons will fill up the lowest $x^2-y^2$ and $xy$ levels, thus giving an orbital singlet. Such a solution was obtained in the previous density-functional calculations.\cite{Rau}
Then the Co/MgO (001) would be a spin-only system, which is however in contradiction with the experiments.\cite{Rau} The reason for this discrepancy could well be that although the $x^2-y^2$ and $xy$ orbitals are two lowest levels in a single electron picture, their double occupation may be energetically unfavorable as their inter-orbital Coulomb interaction is remarkable due to their common in-plane character.\cite{Wu09,Burnus} As a result, the minority-spin $xy$ electron could be prompted to the higher lying $xz$/$yz$ doublet to reduce the inter-orbital Coulomb repulsion, which may over compensate the energy cost associated with the ligand field excitation, see Figure 2(b).

\begin{figure}[t]
\centering \includegraphics[width=7cm]{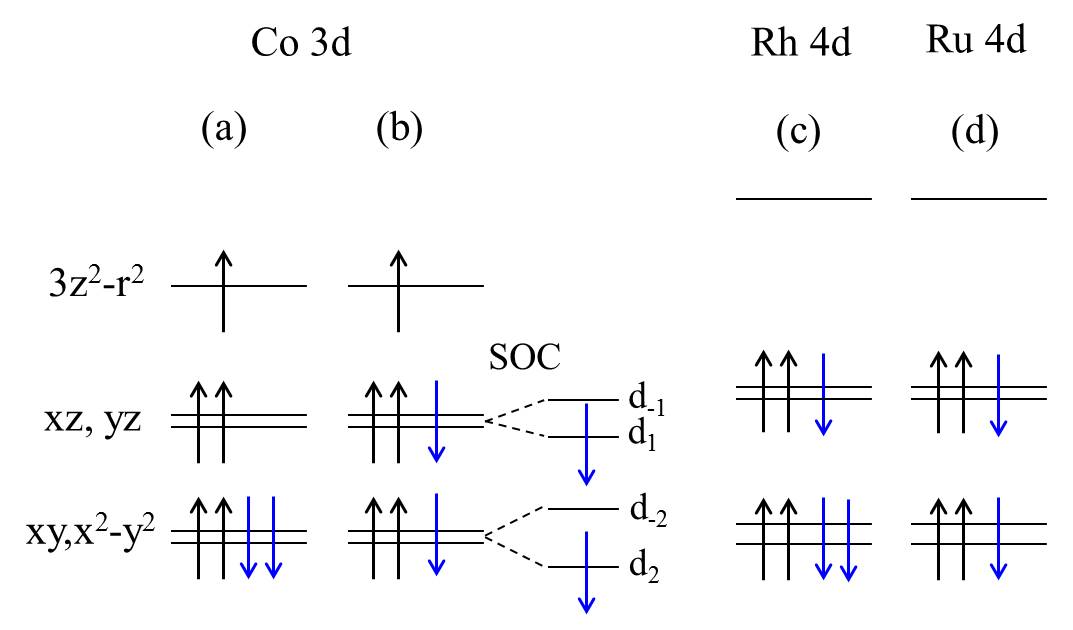}
\caption{
Ligand-field level diagrams of high-spin Co 3d, low-spin Rh 4d and intermediate-spin Ru 4d orbitals. (a) and (b) show the orbital multiplet effect, and the SOC is active in (b), (c), and (d).
}
\end{figure}

To test this scenario, we have done constraint LSDA+U calculations to compare the two above configuration states directly. We set their respective occupation number matrix and thus orbitally dependent potentials, and then do self-consistent calculations including a full electronic relaxation. (Otherwise, some states of the concern or even the ground state may not be achieved.) An advantage of this procedure is such that we can reliably determine the magnetic ground state by a direct comparison of the different states.\cite{Wu09,Wu06}

Figure 3(a) shows the orbitally resolved density of states (DOS) of the HS state with the double occupation of the minority-spin $x^2-y^2$ and $xy$ orbitals (with the majority-spin channel being fully occupied). One can see that the $x^2-y^2$ and $xy$ levels are almost degenerate, which turns out to be crucial for producing a giant MAE as seen below. Note that although the minority-spin $3z^2-r^2$ orbital is formally unoccupied, it has a finite occupation due to the strong pd$\sigma$ covalency between the Co adatom and the underlying O atom. This holds true for all the cases discussed in this paper. Figure 3(b) instead shows the DOS results of the HS state with another double occupation of the minority-spin $x^2-y^2$ and $xz$ (or $yz$). Apparently, our constraint LSDA+U calculations have achieved both the solutions, and from the calculated total energies, we find that the later solution is indeed more stable than the former one by 275 meV/Co, thus verifying the above idea.

\begin{figure}[t]
\centering \includegraphics[width=7cm]{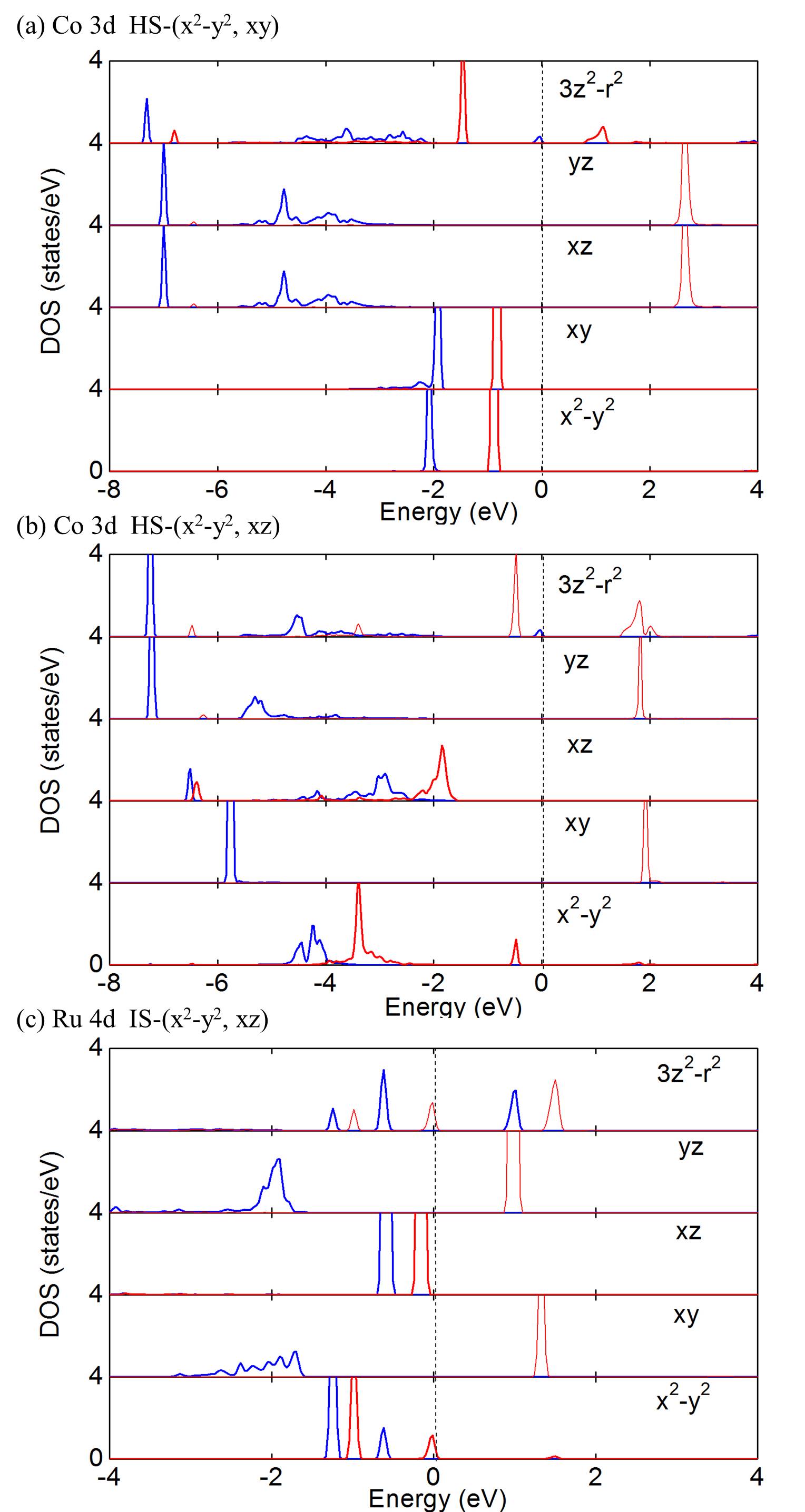}
\caption{(Color online)
Orbitally resolved DOS of HS Co 3d states with (a) the down-spin ($x^2-y^2$, $xy$) or (b) ($x^2-y^2$, $xz$) occupation, and of (c) IS Ru 4d, all calculated by LSDA+U. The blue (red) curves stand for the up (down) spin channel. Fermi level is set at zero energy. See also Figure 2.
}
\end{figure}

Then we arrive at the Co-HS ground-state solution [Figs. 2(b) and 3(b]] with a half filling of both the minority-spin $x^2-y^2$/$xy$ doublet (in a good approximation) and the genuine $xz$/$yz$ doublet. In this case, the SOC is operative, and then the half-filled $x^2-y^2$/$xy$ doublet will produce the occupied $d_2$ ($l_z$ = 2) state in the down-spin channel, and the half-filled $xz$/$yz$ doublet will yield the occupied $d_1$ ($l_z$ = 1) state, thus formally giving a maximal orbital moment of 3 $\mu_B$, see Figure 2(b). We plot in Figure 4(a) the DOS results of the ground state solution given by our LSDA+SOC+U calculations, where one can clearly see the occupied $d_2$ and $d_1$ states in the down-spin channel. This solution gives a local spin moment of 2.44 $\mu_B$ at the HS (S = 3/2) Co site, and a huge orbital moment of 2.48 $\mu_B$ as well, see Table I. Both values are reduced from their ideal ones of 3 $\mu_B$ by the finite p-d covalency.

\begin{figure}[t]
\centering \includegraphics[width=7cm]{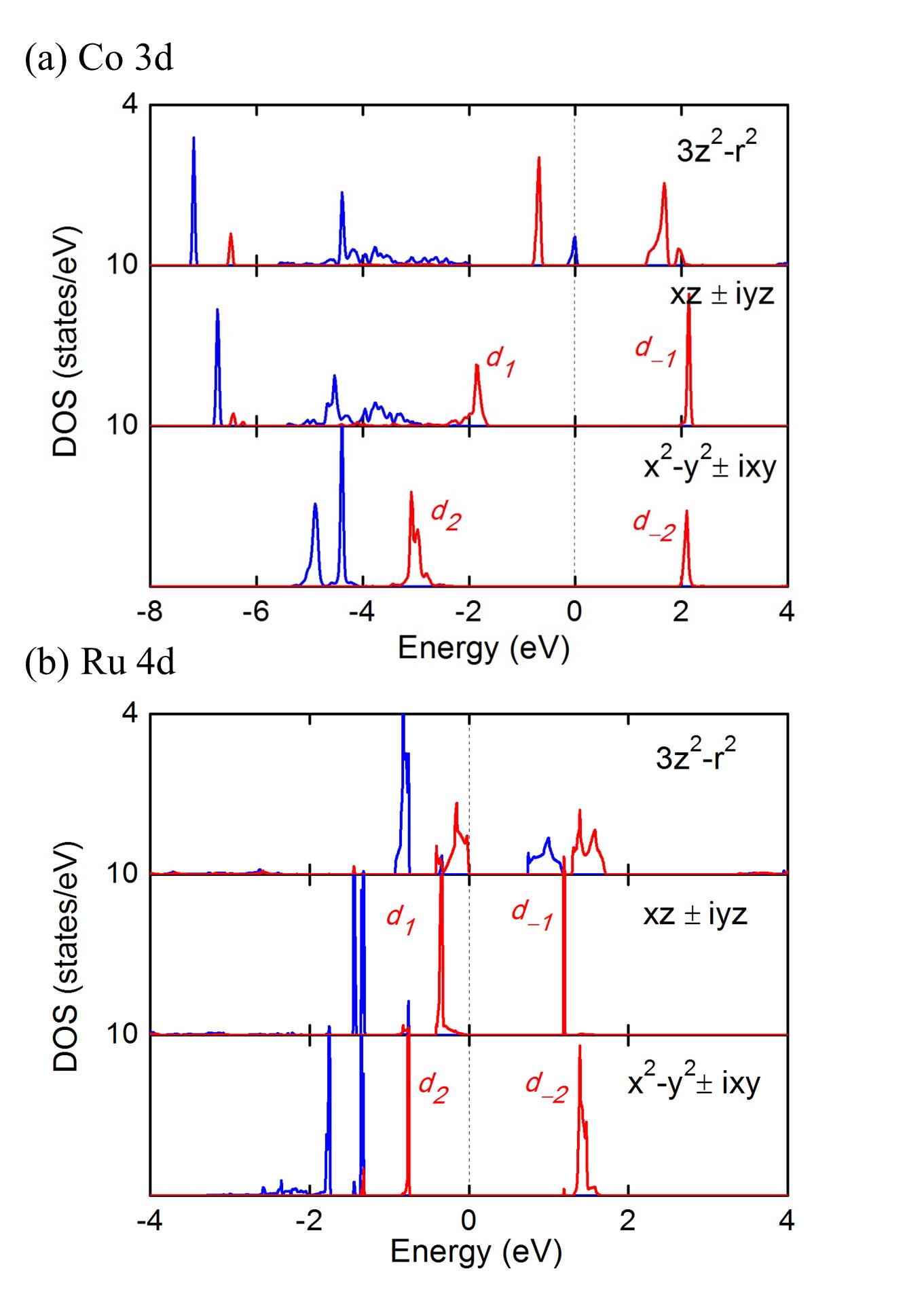}
\caption{(Color online)
The LSDA+SOC+U ground state solution of the HS Co 3d and the IS Ru 4d, both with the down-spin $d_2$ and $d_1$ occupation.
}
\end{figure}

Owing to the huge orbital moment and the remarkable SOC effect, Co/MgO (001) would display a giant MAE. We have assumed in our LSDA+SOC+U calculations two magnetization directions, either along [001] or along [100]. By comparison of these two magnetization directions, our total energy results show that the MAE is up to 90 meV/Co, with the highly preferential easy axis being out of plane, see Table I. Then Co/MgO (001) has both the spin and orbital moments firmly fixed along the surface normal direction. Thus the present results well account for the recent discovery, which reports on the giant MAE of about 60 meV/Co.\cite{Rau,note} Here the physics insight is such that owing to the special ligand field for the HS Co adatoms on top of the surface oxygens of MgO (001), the orbital multiplet effect brings about (near) degenerate levels and then the SOC gives rise to the huge orbital moment and the giant MAE.


As the SOC strength is proportional to $Z^4$, one may assume that heavier Rh or Ir adatoms (in the same group as Co in the Periodic Table) could have an even higher MAE, when they reside at the stable O-top sites. However, as Rh 4d or Ir 5d orbital becomes spatially extended, the ligand field effect gets stronger, mainly due to the enhanced p-d covalency. As a result, the antibonding $3z^2-r^2$ level is pushed too high in the ligand field level diagram (and formally becomes fully unoccupied), see Figure 2(c). Then a HS state cannot be stabilized by the moderate (weak) Hund exchange of the 4d (5d) TM such as Rh (Ir). Instead the stable Rh-LS (low-spin) state as usual ($4d^7$, S=1/2), or the Ir-LS state ($5d^7$, S=1/2), has such a configuration state as plotted in Figure 2(c). Then regardless of whether the minority-spin $xz$/$yz$ doublet or the $x^2-y^2$/$xy$ 'doublet' is half filled due to the orbital multiplet effect, the orbital moment will be largely reduced (compared with the above Co case) by both the decreasing orbital degrees of freedom and the enhanced p-d covalency. Indeed, our LSDA+SOC+U calculations find that the stable Rh-LS ground state gives a spin (orbital) moment of only 0.90 (0.93) $\mu_B$ for each Rh adatom. The corresponding MAE is calculated to be only 8 meV/Rh. The spin and orbital moments should even be smaller for the Ir adatoms, due to the more delocalized character of Ir 5d electrons. Then Rh and Ir adatoms would not be a good option to achieve a larger MAE, compared with the Co adatoms.

Having a look again at Figure 2(c) about the level diagram of the above Rh-LS state, we are optimistic to find a candidate which can have a larger MAE. We just move to the left neighbors in the Periodic Table and now deal with Ru and Os. As Ru (Os) sees a very similar ligand field as Rh (Ir), its one electron less d shell would sustain the Ru intermediate-spin (IS) state ($4d^6$, S=1), see Figure 2(d). Then its electronic configuration is similar to the HS Co [Fig. 2(b)], except for the $3z^2-r^2$ hole in the former. If this is the case, one would readily get an even larger MAE.

Again, our calculations find the most stable O-top sites for the Ru (Os) adatoms on MgO (001) (see Table 1), being the same as the Co adatoms. Moreover, we have compared the IS state with a possible HS or LS state through a set of constraint LSDA+U calculations, and we have indeed got the Ru-IS ground state [Figure 3(c)], which well represents the level diagram plotted in Figure 2(d). (In addition, the strong pd  covalency makes the otherwise fully unoccupied $3z^2-r^2$ orbital partially occupied.) Then the SOC is active [see Figs. 2(b) and 2(d)], and the LSDA+SOC+U calculations are performed for the IS ground state. Figure 4(b) shows the calculated Ru 4d DOS results, and the down-spin $d_2$ and $d_1$ occupations are evidenced, giving again a huge orbital moment. In this IS ground state solution, the Ru local spin moment is calculated to be 1.73 $\mu_B$ (see Table I), and the orbital moment is reduced to 1.81 $\mu_B$, due to the strong covalency. Owing to the big orbital moment and the significant SOC, here the MAE is calculated to be 110 meV/Ru (with the easy axis being again out of plane), even higher than the above 90 meV/Co. Similarly, the IS-Os adatoms on MgO (001) could have a highest MAE, as the SOC strength is about 0.4-0.5 eV for Os 5d electrons (being about 3 times as high as Ru 4d SOC of about 0.15 eV). Indeed, our LSDA+SOC+U calculations for Os/MgO (001) find the giant MAE is up to 208 meV/Os, see the results in Table I. Therefore, we propose that Ru and Os adatoms on MgO (001) would have the giant MAE, which calls for an experimental verification.


In conclusion, in this work a giant MAE is explored for TM adatoms on MgO (001) surface, using ligand field analysis and constraint LSDA+U and LSDA+SOC+U calculations. We find that at the most stable O-top adsorption site, the TM adatoms (Co, Ru, and Os) see a special tetragonal ligand field. The orbital multiplet effect of the HS Co ($3d^7$, S=3/2) and of the IS Ru and Os ($4d^6$/Ru and $5d^6$/Os, both S=1) makes their significant SOC active, thus producing a huge orbital moment (formally with $l_z$ = 3) and a giant MAE. The present results well explain the recent discovery\cite{Rau} and predict Ru and Os adatoms on MgO (001) to have the giant MAE. Then the giant magnetic anisotropy could be engineered at the atomic scale. This prediction deserves experimental tests and may also have an implication for the nanoscale information storage.\\

This work was supported by the NSF of China (Grant Nos. 11274070 and 11474059), MOE Grant No. 20120071110006, and ShuGuang Program of Shanghai (Grant No. 12SG06).


\begin{thebibliography}{22}
\bibitem{note}
Actually, a zero-field splitting of about 60 meV was observed in [11], which
corresponds to the SOC energy $\lambda$$L\approx$ 60 meV. Then the MAE could be up to 90 meV for the $S$ = 3/2 Co adatom.
\end{thebibliography}
\end{document}